\documentclass[12pt,twocolumn]{aastex63}

\newcommand{\kms}{\rm km~s^{-1}}
\newcommand{\kmsmpc}{\rm km~s^{-1}~Mpc^{-1}}

\usepackage{color}
\usepackage{multirow}
\usepackage[ddmmyyyy]{datetime}

\begin{document}

\title{Co-evolution of the Brightest Cluster Galaxies and their Host Clusters in Illustris-TNG}

\author{Jubee Sohn}
\affiliation{Smithsonian Astrophysical Observatory, 60 Garden Street, Cambridge, MA 02138, USA}

\author{Margaret J. Geller}
\affiliation{Smithsonian Astrophysical Observatory, 60 Garden Street, Cambridge, MA 02138, USA}

\author{Mark Vogelsberger}
\affiliation{Department of Physics, Kavli Institute for Astrophysics and Space Research, Massachusetts Institute of Technology, Cambridge, MA 02139, USA}

\author{Ivana Damjanov}
\affiliation{Department of Astronomy and Physics, Saint Mary's University, 923 Robie Street, Halifax, NS B3H 3C3, Canada}
\affiliation{Canada Research Chair in Astronomy and Astrophysics, Tier II}

\email{jubee.sohn@cfa.harvard.edu}

\begin{abstract}
We use the Illustris-TNG simulations to explore the dynamic scaling relation between massive clusters and their central galaxies (BCGs). The Illustris TNG300-1 simulation we use includes 280 massive clusters with $M_{200} > 10^{14}$ M$_{\odot}$ enabling a robust statistical analysis. We derive the line-of-sight velocity dispersion of the stellar particles of the BCGs ($\sigma_{*, BCG}$), analogous to the observed BCG stellar velocity dispersion. We also compute the subhalo velocity dispersion to measure the cluster velocity dispersion ($\sigma_{cl}$). Both $\sigma_{*, BCG}$ and $\sigma_{cl}$ are proportional to the cluster halo mass, but the slopes differ slightly. Thus like the observed relation, $\sigma_{*, BCG} / \sigma_{cl}$ declines as a function of $\sigma_{cl}$, but the scatter is large. We explore the redshift evolution of $\sigma_{*, BCG} - \sigma_{cl}$ scaling relation for $z \lesssim 1$ in a way that can be compared directly with observations. The scaling relation has a similar slope at high redshift, but the scatter increases because of the large scatter in $\sigma_{*, BCG}$. The simulations imply that high redshift BCGs are dynamically more complex than their low redshift counterparts. 
\end{abstract}

\section{INTRODUCTION}

Brightest cluster galaxies (BCGs) are a special population of luminous, massive galaxies. BCGs are usually found at the bottom of the cluster potential well indicated by the peak X-ray emission (e.g., \citealp{Jones84, Postman95, Lin04, Sanderson09, Lauer14, Lopes18}). This coincidence indicates that the formation of the BCGs is tightly associated with the formation of cluster halos.

Hierarchical structure formation models suggest that massive clusters form and evolve through stochastic accretion of surrounding material (e.g., \citealp{vandenBosch02, McBride09, Zhao09, Fakhouri10, Kravtsov12, Haines18}). BCGs in the cluster grow through active accretion of other cluster members and material stripped from other galaxies. BCG evolution is more complex than cluster evolution because baryonic physics plays an important role. 

Comparison between the BCG and cluster mass tests the co-evolution of a cluster and its BCG (e.g., \citealp{Lin04, OlivaAltamirano14, Lin17, Kravtsov18, Wen18, Erfanianfar19, GoldenMarx21}). Many studies derive the ratio between the stellar mass of a central galaxy and its host halo mass. The ratio between the stellar mass of a central galaxy and its host halo mass has a peak at $M_{halo} \sim 10^{12}$ M$_{\odot}$ (e.g., \citealp{Conroy06, Behroozi10, Behroozi19, Guo10, Moster10}). In other words, the mass ratio declines as a function of halo mass within the cluster mass range. This decline suggests that the stellar mass growth of BCGs in massive halos is suppressed by strong feedback processes including active galactic nuclei (AGN) (e.g., \citealp{DiMatteo05, McNamara07, Kravtsov12, Weinberger17}). 

\citet{Sohn20a} explore the relation between the BCGs and their host clusters based on dynamical properties measured from observations. They use the central stellar velocity dispersion ($\sigma_{*, BCG}$) of the BCG which probes the BCG subhalo mass. The central stellar velocity dispersion is proportional to the dark matter velocity dispersion, which is proportional to the dark matter halo mass \citep{Zahid18}. \citet{Sohn20a} also compute the cluster velocity dispersion ($\sigma_{cl})$ that probes the cluster mass. These dynamical properties are powerful tools because they are insensitive to the complex baryonic physics. 

\citet{Sohn20a} show that $\sigma_{*, BCG} / \sigma_{cl}$ decreases as a function of $\sigma_{cl}$ based on the HeCS-omnibus sample that compiles dense spectroscopy of 227 clusters. A similar relation appears in other cluster samples (e.g., \citealp{Kim17, Sohn21b}). The decreasing $\sigma_{*, BCG} / \sigma_{cl}$ ratio at higher $\sigma_{cl}$ indicates that the mass fraction associated with the BCG halo decreases in higher mass clusters. \citet{Sohn20a} and \citet{Sohn21b} show that BCG growth in more massive clusters slows down because interaction between the BCG and  other cluster members is suppressed by the large cluster velocity dispersion. BCGs in less massive clusters, where the cluster dispersion is comparable with the BCG central stellar velocity dispersion, can continue to grow.  

\citet{Sohn20a} and \citet{Sohn21b} compare the observed $\sigma_{*, BCG} - \sigma_{cl}$ relation with theoretical relations from \citet{Dolag10} and \citet{Remus17}. These theoretical relations predict that the $\sigma_{*, BCG} / \sigma_{cl}$ relation is constant over a large $\sigma_{cl}$ range. \citet{Marini21} revisit this issue using the DIANOGA hydrodynamic zoom-in simulations. \citet{Marini21} show that $\sigma_{*, BCG} / \sigma_{cl}$ declines slightly as a function of $\sigma_{cl}$; this revised theoretical relation is consistent with the observations. 

Here we use the Illustris-TNG 300-1 cosmological hydrodynamic simulation \citep{Springel18} to investigate the $\sigma_{*, BCG} - \sigma_{cl}$ relation. Based on the 280 massive clusters ($M_{200} > 10^{14}$ M$_{\odot}$) in the Illustris-TNG simulation TNG300-1, we carry out a statistical exploration of clusters and their BCGs. For comparison with the observed data, we derive the $\sigma_{cl}$ and $\sigma_{*, BCG}$ from Illustris-TNG in analogy with the observed properties. We also investigate the redshift evolution of $\sigma_{*, BCG} - \sigma_{cl}$ relation for $z \lesssim 1$ providing an important baseline for future cluster observations. 

We describe the Illustris-TNG simulation and the techniques we use for deriving $\sigma_{*, BCG}$ and $\sigma_{cl}$ in Section \ref{sec:data}. We demonstrate the simulated $\sigma_{BCG} - \sigma_{cl}$ relation in Section \ref{sec:result}. In Section \ref{sec:discuss}, we compare the simulated and observed scaling relations, and the redshift evolution of the scaling relations. We conclude in Section \ref{sec:conclusion}. We adopt the Planck cosmological parameters  \citep{Planck16} with $H_{0} = 67.74~\kmsmpc$, $\Omega_{m} = 0.3089$, $\Omega_{\Lambda} = 0.6911$.

\section{DATA}\label{sec:data}
\subsection{The Illustris-TNG Simulation}

Illustris-TNG is a set of cosmological magnetohydrodynamic (MHD) simulations of galaxy formation \citep{Springel18}. Illustris-TNG improves on its predecessor, Illustris \citep{Vogelsberger14a,Vogelsberger14b}, by extending the mass range of simulated halos by simulating larger volumes. Illustris-TNG also includes an improved galaxy formation model that implements black hole (BH) driven wind feedback that affects the highest mass galaxies \citep{Weinberger17, Pillepich18a, Pillepich18b}. 

We use the TNG300-1 simulation selected from the set of Illustris-TNG simulations. TNG300 is one of the largest cosmological simulations covering a $\sim 300$ Mpc cube \citep{Vogelsberger20}. The TNG300 set includes three simulations with different mass resolution. We use the TNG300-1, the highest resolution simulation in this box size, with dark matter particle mass $m_{DM} = 59 \times 10^{6}$ M$_{\odot}$ and target gas cell mass $m_{baryon} = 11 \times 10^{6}$ M$_{\odot}$. 

The large volume of TNG300-1 enables study of clusters and their BCGs based on a large number of simulated clusters. TNG300-1 includes 280 massive clusters with $M_{200} > 10^{14}$ M$_{\odot}$, five times larger than previous simulated cluster samples (e.g., \citealp{Dolag10, Bahe17, Marini21}). The mass resolution of TNG300-1 is also better than in previous simulations \citep{Vogelsberger20}. Our analysis extends and complements work by \citet{Marini21} who explore the relation between the velocity dispersion of BCGs and their clusters with the DIAGONA set of simulations. These simulations provide a set of 57 simulated cluster halos with total dark matter mass larger than $7 \times 10^{9}$ M$_{\odot}$, a much smaller sample than we  obtain based on Illustris-TNG.

\subsection{Cluster Velocity Dispersion}

We use a group catalog derived from the TNG300-1 simulation to select cluster-like halos. TNG300-1 provides a group catalog constructed by applying a standard friends-of-friends algorithm with a fractional linking length $b = 0.2$ times the mean separation of galaxies at a given redshift. This group catalog lists the group properties including the critical mass and size of the groups. Here, we obtain $M_{200}$ (and $R_{200}$), the total mass of the group enclosed within a sphere of mean density 200 times the critical density at the group redshift (in this case, $z = 0.0$). 

We select 280 massive group halos with $M_{200} > 10^{14}$ M$_{\odot}$ (hereafter cluster halos). We apply this mass limit for comparison with observed clusters (see Section \ref{sec:obs}). This mass limit is comparable with the mass limit of other widely used cluster catalogs (e.g., the redMaPPer, \citealp{Rykoff14, Rykoff16}).

The group catalog also lists the properties of subhalos belonging to each group halo. We use the position and velocity of the subhalos in the 280 massive cluster halos to compute the cluster velocity dispersion. We calculate the projected distance of the subhalos from the cluster halo center:
\begin{equation}
R_{cl} = \sqrt{\Delta X^{2} + \Delta Y^{2}},
\end{equation}
where $\Delta X$ and $\Delta Y$ are the separation between the subhalo and central cluster halo positions. 

We compute the velocity dispersion among the subhalos. This subhalo velocity dispersion corresponds to the observed cluster velocity dispersion. Observationally, the cluster velocity dispersion is the velocity dispersion of cluster member galaxies (e.g., \citealp{Sohn20a}). 

We also restrict the analysis to  subhalos with stellar mass larger than $10^{9}$ M$_{\odot}$, roughly corresponding to the stellar mass limit of dense spectroscopic surveys of massive clusters \citep{Sohn17}. For direct comparison with the observed cluster velocity dispersion in \citet{Sohn20a}, we derive the cluster velocity dispersion based on subhalos within $R_{cl} < R_{200}$. We then compute the cluster velocity dispersion based on 27 to 821 subhalos (with a median 75). In analogy with the observed cluster velocity dispersion, we use the bi-weight technique \citep{Beers90} to compute the velocity dispersions. Hereafter, we refer to this measure as the cluster velocity dispersion ($\sigma_{cl}$). We compute the cluster velocity dispersion uncertainty, the $1\sigma$ standard deviation, from 1,000 bootstrap resamplings. 

We note that our approach differs from previous works using numerical simulations (e.g., \citealp{Dolag10, Remus17, Marini21}). Previous work measures the velocity dispersion of intracluster particles \citep{Dolag10, Remus17} or of the dark matter particles belonging to the cluster halo \citep{Marini21}. Our approach of measuring the subhalo velocity dispersion provides a more direct comparison with the observed cluster velocity dispersion.

Because the simulation provides the three-dimensional distribution of positions and velocities of subhalos, we can derive the 3D velocity dispersion. However, the 3D velocity dispersion is not observable. Thus, we derive the observable line-of-sight velocity dispersion of subhalos based on the relative velocity differences between subhalos and the cluster halo center in the $z-$direction.

\begin{figure}
\centering
\includegraphics[scale=0.28]{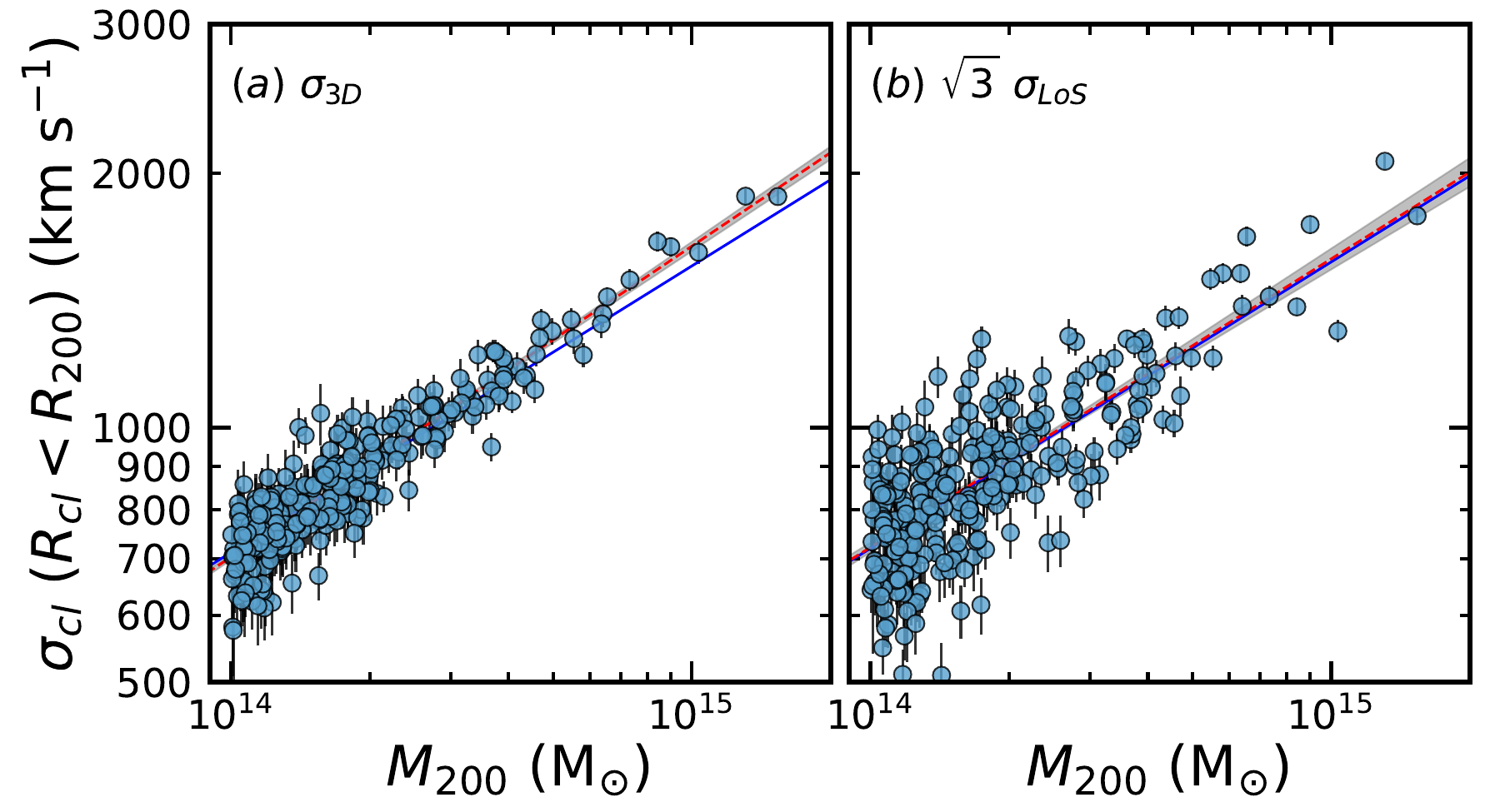}
\caption{(a) The 3D velocity dispersion of cluster halos as a function of cluster mass ($M_{200}$). (b) Same as panel (a), but for the line-of-sight velocity dispersion times $\sqrt{3}$. Gray shaded regions mark the $1\sigma$ distribution of the MCMC fits. Red lines show the best-fit power laws with slopes of $\alpha = 0.368 \pm 0.008$ and $0.341 \pm 0.016$, respectively. Blue lines are the best-fit relation ($\alpha = 0.339$) derived from \citet{Marini21}. }
\label{sigma_m200}
\end{figure}

Figure \ref{sigma_m200} shows the cluster velocity dispersion as a function of $M_{200}$. Figure \ref{sigma_m200} (a) shows the 3D velocity dispersion and Figure \ref{sigma_m200} (b) displays the line-of-sight velocity dispersion (times $\sqrt{3}$). Both velocity dispersions are tightly correlated with $M_{200}$.

We derive the best-fit power-law relation $\sigma_{cl} \propto M_{200}^{\alpha}$ following \citet{Marini21}. We use a Markov chain Monte Carlo (MCMC) technique to derive the best-fit relations. The best-fit relation from the massive TNG halos has a slope $\alpha = 0.368 \pm 0.008$ for the 3D velocity dispersion and $\alpha = 0.343 \pm 0.016$ for the LoS velocity dispersion. These slopes are consistent with the slope derived from the other simulations (e.g., $\alpha = 0.339$, \citealp{Marini21}). We also note that the 3D velocity dispersion is consistent with $\sigma_{cl, LoS} \times \sqrt{3}$. 

\subsection{Velocity Dispersions of Brightest Cluster Galaxies}

The group catalog from TNG300-1  lists the most massive subhalo in each group. We assume that the most massive subhalos correspond to the brightest cluster galaxies (BCGs) in the observations. Hereafter, we refer to the most massive subhalo in each of the 280 massive clusters as the BCG. 

TNG300-1 provides the properties of stellar particles that belong to each subhalo, which is identified by the SUBFIND algorithm. We select stellar particles within each of the 280 BCGs to compute the BCG velocity dispersion. We use the stellar particles within cylindrical volumes with three different apertures: (a) $R_{proj} < 3$ kpc, (b) $< 50$ kpc, and (c) $< R_{h}$, where $R_{proj} = \sqrt{\Delta X_{*}^2 + \Delta Y_{*}^2}$, and $\Delta X_{*}$ and $\Delta Y_{*}$ are the distance between stellar particles and the BCG subhalo center along the $x-$ and $y-$axes. $R_{h}$ is the comoving radius that contains half of the stellar mass of the subhalo. Thanks to the high resolution of TNG300-1, the BCGs in our sample include a large number of particles. For example, the BCG in the least massive cluster halo consists of 3855 stellar particles within a cylindrical volume with a 3 kpc projected radius. Therefore, the BCG stellar velocity dispersion is insensitive to small number statistics. In analogy with the observed BCG stellar velocity dispersion, we compute the line-of-sight velocity dispersions of the stellar particles. We use the bi-weight technique to compute the BCG velocity dispersion. The BCG velocity dispersion uncertainty from 1000 bootstrap resamplings is tiny ($< 3~\kms$). 

\begin{figure*}
\centering
\includegraphics[scale=0.38]{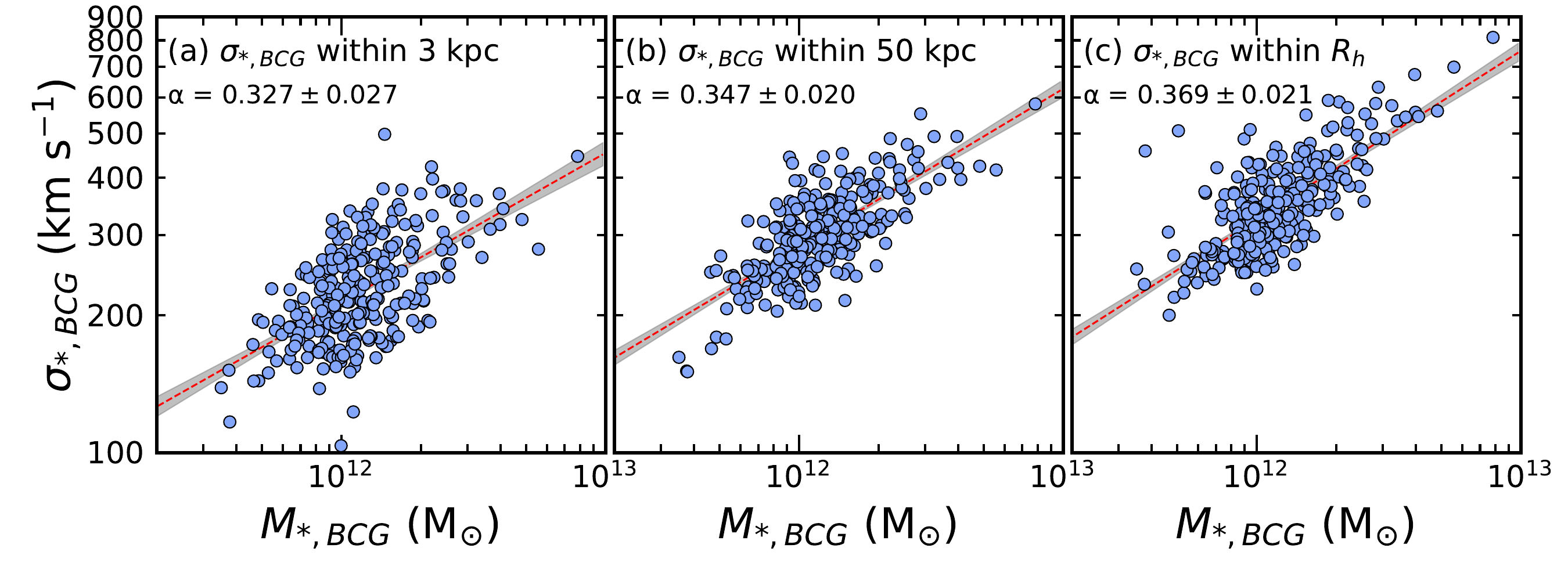}
\caption{The line-of-sight stellar velocity dispersion of BCGs measured within (left) 3 kpc, (middle) 50 kpc, and (right) the half-mass radius as a function of the stellar mass of the BCGs. Red dashed lines show the best-fit relation. Gray shaded regions mark the $1\sigma$ distribution of the MCMC fits. }
\label{bcg_sigma_aperture}
\end{figure*}

Figure \ref{bcg_sigma_aperture} shows the stellar velocity dispersion of the BCGs as a function of the stellar mass of the BCGs. Here, the stellar mass of the BCGs is the sum of masses of the stellar particles within the radius $V_{max}$ ({\it SubhaloMassInRadType} in the TNG300-1 catalog). 

The stellar velocity dispersion of the BCGs is correlated with the BCG subhalo mass. Red lines in Figure \ref{bcg_sigma_aperture} show the best-fit power law ($\sigma_{*, BCG} \propto M_{200}^{\alpha}$) based on the MCMC technique; gray shaded regions show $1\sigma$ distribution of the MCMC fits. The slopes of the relations vary from 0.327 to 0.369 depending on the aperture. Interestingly, the BCG velocity dispersion measured within a larger radius is generally more tightly correlated with the BCG mass. 

The observed cluster sample is based on BCG velocity dispersions measured within a fiducial 3 kpc aperture. Therefore, using the BCG subhalo velocity dispersion measured within 3 kpc enables direct comparison with the observations. 

\begin{figure}
\centering
\includegraphics[scale=0.28]{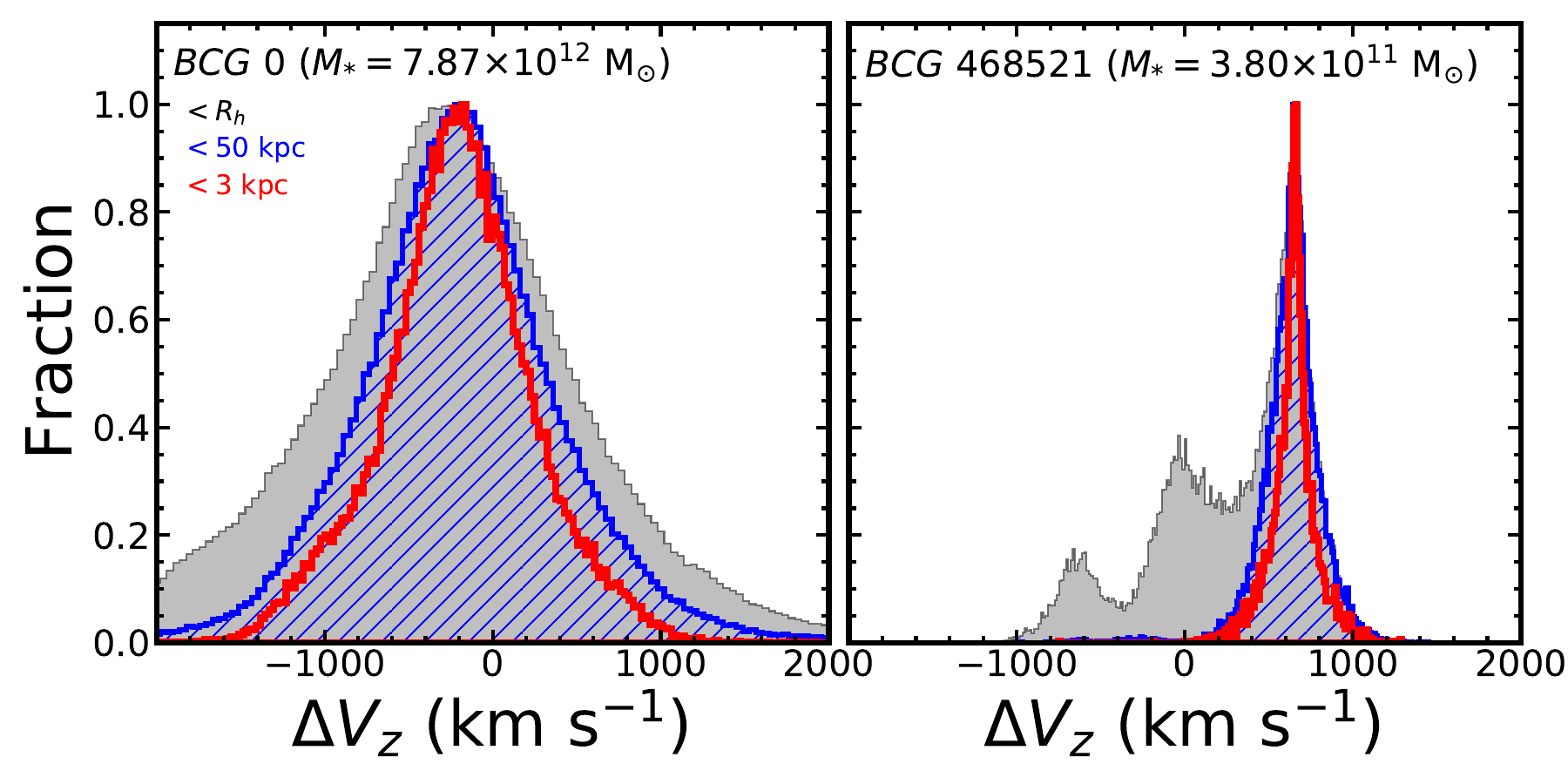}
\caption{The line-of-sight velocity distribution of stellar particles within 3 kpc (the red open histogram), 50 kpc (the blue hatched histogram), and the half-mass radius (the gray filled histogram) in two BCG halos in our sample. The target BCGs are located in the most massive (left) and the least massive clusters (right) in our sample. }
\label{bcg_los_vdisp}
\end{figure}

\begin{figure}
\centering
\includegraphics[scale=0.19]{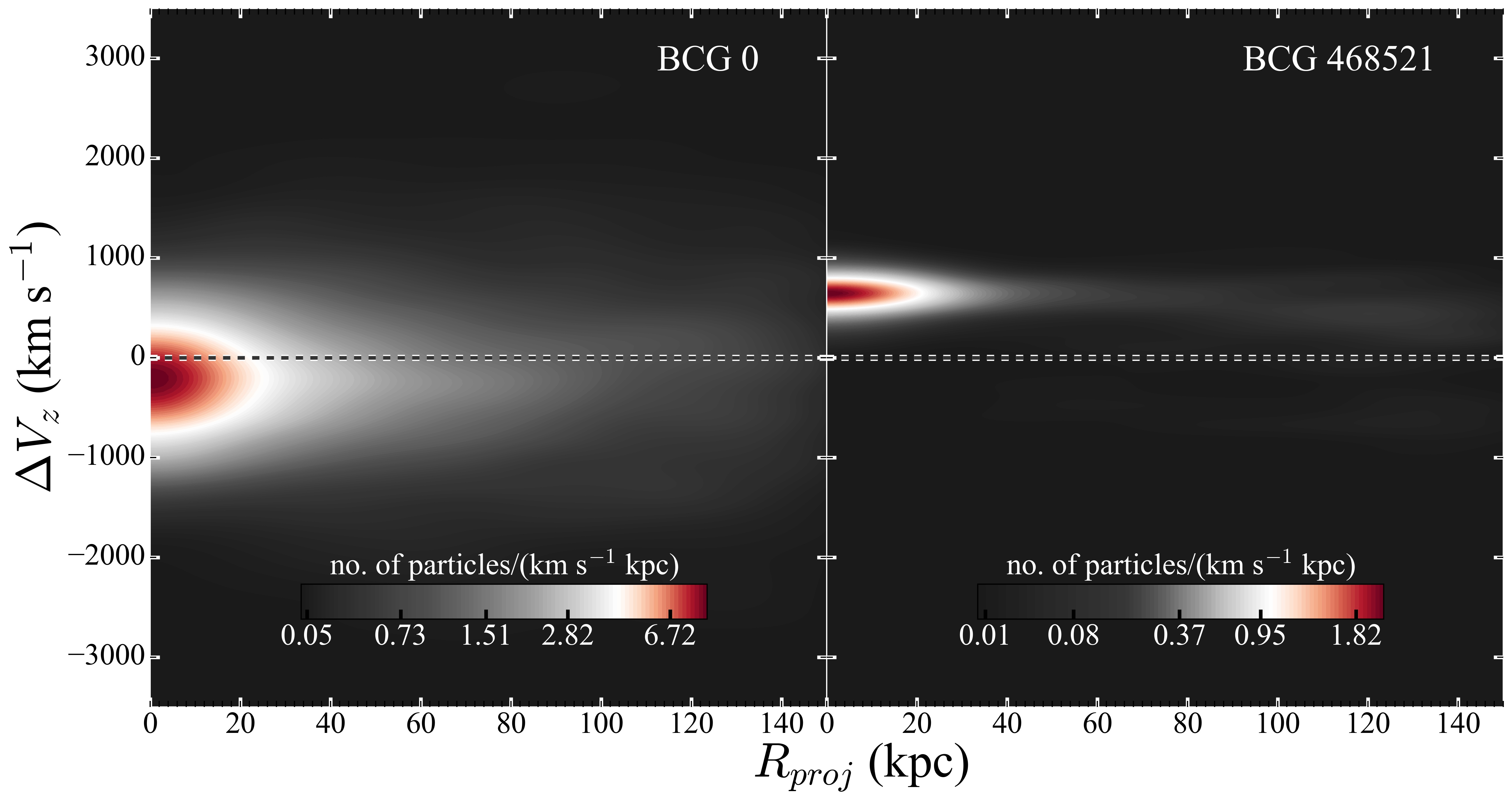}
\caption{The line-of-sight velocity as a function of projected distance (phase space diagram) from the subhalo center of stellar particles within the two BCG subhalos (the same BCGs in Figure \ref{bcg_los_vdisp}). }
\label{bcg_rv}
\end{figure}

Figure \ref{bcg_los_vdisp} and Figure \ref{bcg_rv} demonstrate another reason for using the velocity dispersion measured within a smaller aperture. For this demonstration, we select two BCGs in the most massive and the least massive cluster halos in our sample. 

In Figure \ref{bcg_los_vdisp}, we show the line-of-sight velocity distribution of the stellar particles within 3 kpc (the red open histogram), 50 kpc (the blue hatched histogram), and $R_{h}$ (the black filled histogram) of the two BCGs. In Figure \ref{bcg_rv}, we plot the phase-space diagram (often called $R-v$ diagram) of the stellar particles of the BCGs; it shows the relative velocity difference of the stellar particles with respect to the BCG center as a function of projected distance from the BCG center. 

In numerical simulations, separating the BCG subhalo from the entire cluster halo is not trivial. For example, \citet{Dolag10} show that stellar particles in the BCG subhalo in numerical simulation consist of two components (see also \citealp{Marini21}); one population is governed by the entire cluster potential and the other component is confined within the BCG subhalo. They derive the velocity dispersion of the two components and interpret the small and large velocity dispersions they derive as representing the BCG and the intracluster velocity dispersions, respectively.

In the most massive cluster, the stellar particles of the BCGs are extended in the line-of-sight velocity direction at larger radius (the left panels of Figure \ref{bcg_los_vdisp} and Figure \ref{bcg_rv}). The stellar particles extended along the line-of-sight may belong to the extended cluster halo as suggested in \citet{Dolag10} (and also in \citealp{Remus17, Marini21}). Using the velocity dispersion within a smaller aperture where the density contrast is high reduces contamination by the much lower density intracluster stellar component. 

For the BCG in a less massive cluster (the right panels of Figure \ref{bcg_los_vdisp} and Figure \ref{bcg_rv}), the situation is more dramatic. In this BCG, the stellar particles associated with the BCG subhalo have multiple components; presumably this BCG is experiencing ongoing interactions. The disturbed stellar components appear at $R_{proj} \gtrsim 30$ kpc, and they impact the velocity dispersion measurements. The velocity dispersion measured within 3 kpc is relatively insensitive to the ongoing activity. In other words, velocity dispersion measurement within 3 kpc is robust. We therefore use the BCG velocity dispersion measured within 3 kpc (hereafter $\sigma_{*, BCG}$). 

\subsection{Comparison Sample from Observations}\label{sec:obs}

Our goal is to compare the $\sigma_{*, BCG} - \sigma_{cl}$ relation from TNG300-1 with the observed relation. For comparison, we use the HeCS-omnibus spectroscopic data compilation \citep{Sohn20a} used for deriving the observed $\sigma_{*, BCG} - \sigma_{cl}$ relation. 

HeCS-omnibus includes 227 massive clusters with extensive spectroscopy. The spectroscopy comes from the SDSS Data Release 16 \citep{SDSS16} along with data collected from various MMT/Hectospec surveys including CIRS \citep{Rines06}, HeCS \citep{Rines13}, HeCS-SZ \citep{Rines16}, HeCS-red \citep{Rines18}, and ACReS \citep{Haines13}. Based on large spectroscopic samples, \citet{Sohn20a} derive the physical properties of the clusters. First, they determine the spectroscopic membership of each cluster based on the caustic technique \citep{Diaferio97, Diaferio99, Serra13}. The caustic technique yields an estimate of the characteristic cluster mass (i.e., $M_{200}$). \citet{Sohn20a} also compute the velocity dispersion for cluster members within $R_{200}$ using the bi-weight technique. This velocity dispersion corresponds to the subhalo velocity dispersion of simulated cluster halos. 

The HeCS-omnibus catalog also provides the physical properties of cluster members including their stellar masses and central stellar velocity dispersions. \citet{Sohn20a} describe the details of these measurements. Here, we briefly introduce the stellar velocity dispersion measurements. For the majority (84/99) of the HeCS-omnibus BCGs, we obtain the stellar velocity dispersion from the Portsmouth reduction \citep{Thomas13} based on the Penalized Pixel-Fitting (pPXF) code \citep{Cappellari04}. There are 15 BCGs with MMT/Hectospec spectroscopy. For these BCGs, we use the University of Lyon Spectroscopic analysis Software (ULySS, \citealp{Koleva09}) to derive the velocity dispersion by comparing the observed spectra with synthetic stellar population templates. Because of the large redshift range, the physical area covered by SDSS/Hectospec fibers vary for the BCGs. We apply an aperture correction (see \citealp{Zahid16, Sohn17}) to obtain the stellar velocity dispersion measured within a fiducial radius of 3 kpc. The aperture correction for the velocity dispersion is negligible ($\sim 3\%$). 

\begin{figure}
\centering
\includegraphics[scale=0.42]{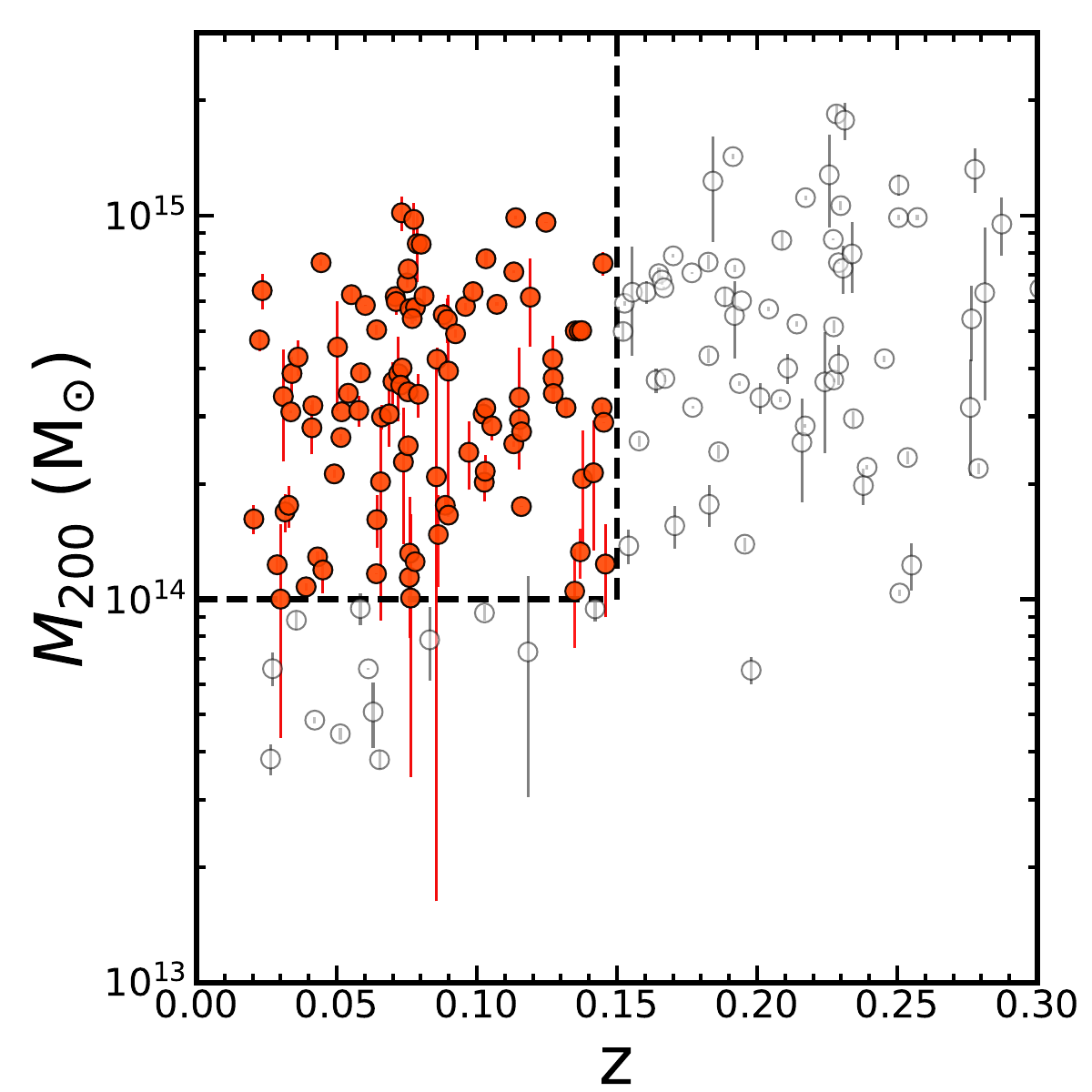}
\caption{$M_{200}$ of HeCS-omnibus clusters (gray circles) as a function of redshift. Red filled circles show 99 clusters with within $z < 0.15$ and $M_{200} > 10^{14}$ M$_{\odot}$ (the dashed lines), the comparison sample for TNG300-1. }
\label{hecs_sample}
\end{figure}

Figure \ref{hecs_sample} shows $M_{200}$ of the HeCS-omnibus clusters as a function of cluster redshift. Most of HeCS-omnibus clusters have masses larger than $10^{14}$ M$_{\odot}$, comparable with mass limit of the other widely used cluster catalogs (e.g., the redMaPPer). We select HeCS-omnibus clusters with $M_{200} > 10^{14}$ M$_{\odot}$ for direct comparison with halos in TNG300-1. We additionally apply a redshift selection ($z < 0.15$) because the HeCS-omnibus sample only includes the few most massive clusters at higher redshift. Furthermore, sampling clusters in a narrow redshift range is relatively insensitive to cluster evolution. The final comparison sample includes 99 HeCS-omnibus clusters with a median redshift of 0.08. We compare this set of observed clusters with the simulated clusters at $z = 0.0$ assuming that evolution over this redshift range is negligible. 

\section{RESULTS}\label{sec:result}

\begin{figure}
\centering
\includegraphics[scale=0.42]{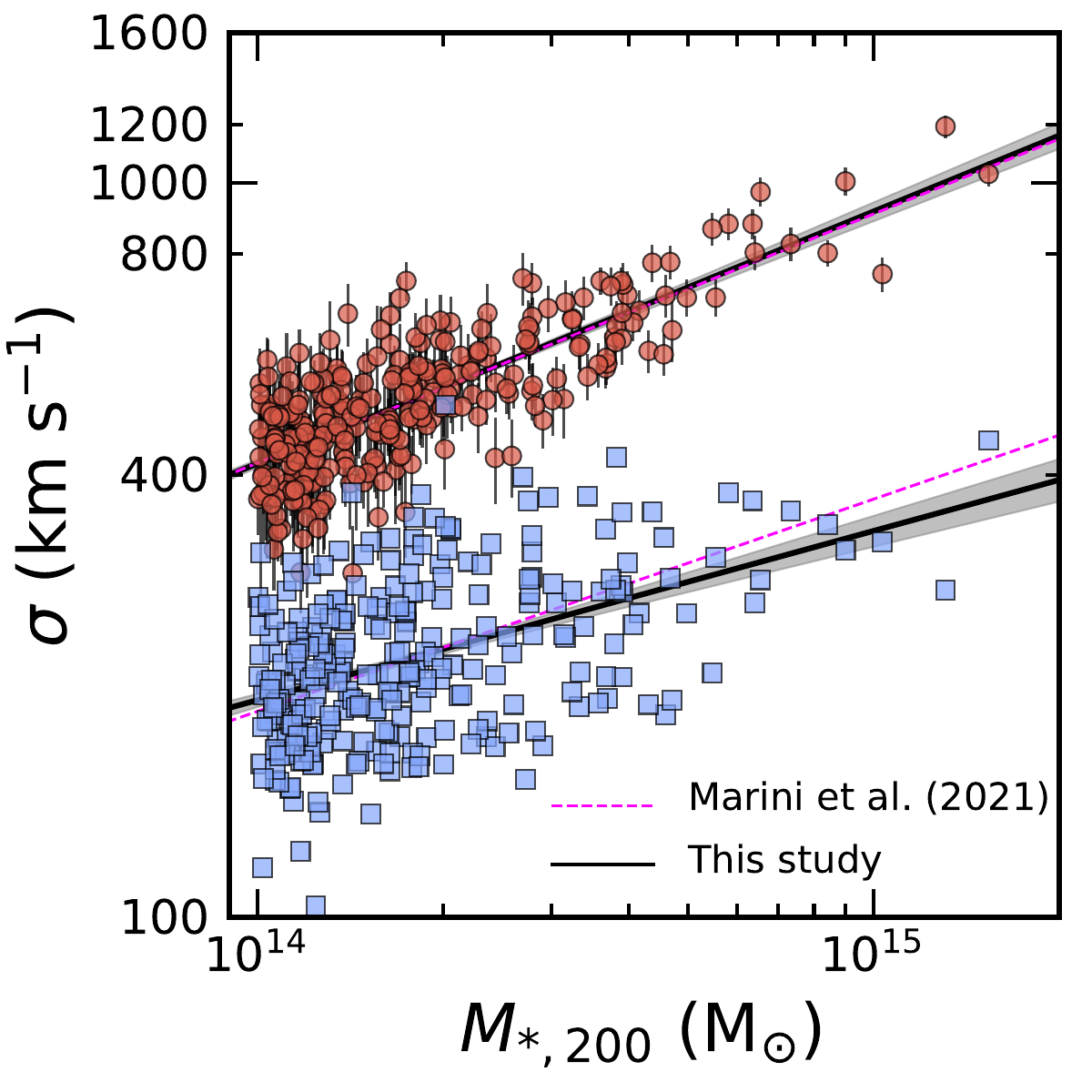}
\caption{The line-of-sight velocity dispersions of clusters (red circles) and their BCGs (blue squares) as a function of cluster mass ($M_{200}$). Black lines show the best-fit relations. Gray shaded regions mark the $1\sigma$ distributions of the MCMC fits. Magenta lines show the best-fit relations from \citet{Marini21}. }
\label{sigma_sigma_m200}
\end{figure}

Figure \ref{sigma_sigma_m200} displays the velocity dispersions of clusters and their BCGs as a function of the mass of the cluster. Similar to Figure \ref{sigma_m200}, we show the line-of-sight velocity dispersions as a function of $M_{200}$. Both $\sigma_{cl}$ and $\sigma_{*, BCG}$ are correlated with $M_{200}$. We derive the best-fit relations:
\begin{equation}
\log \sigma_{cl, R_{cl} < R_{200}} = (0.343 \pm 0.016) \log M_{200} + (-2.185 \pm 0.235), 
\end{equation}
and 
\begin{equation}
\log \sigma_{*, BCG, 3 {\rm kpc}} = (0.230 \pm 0.027) \log M_{200} + (-0.930 \pm 0.384), 
\end{equation}
respectively. Interestingly, the $\sigma_{*, BCG} - M_{200}$ relation is shallower than the $\sigma_{cl} - M_{200}$ relation. This difference in slope is qualitatively consistent with the recent result of \citet{Marini21}. However, the slope of the $\sigma_{*, BCG} - M_{200}$ relation in our sample is shallower than the relation (slope $= 0.289$) derived in \citet{Marini21}. 

\begin{figure}
\centering
\includegraphics[scale=0.42]{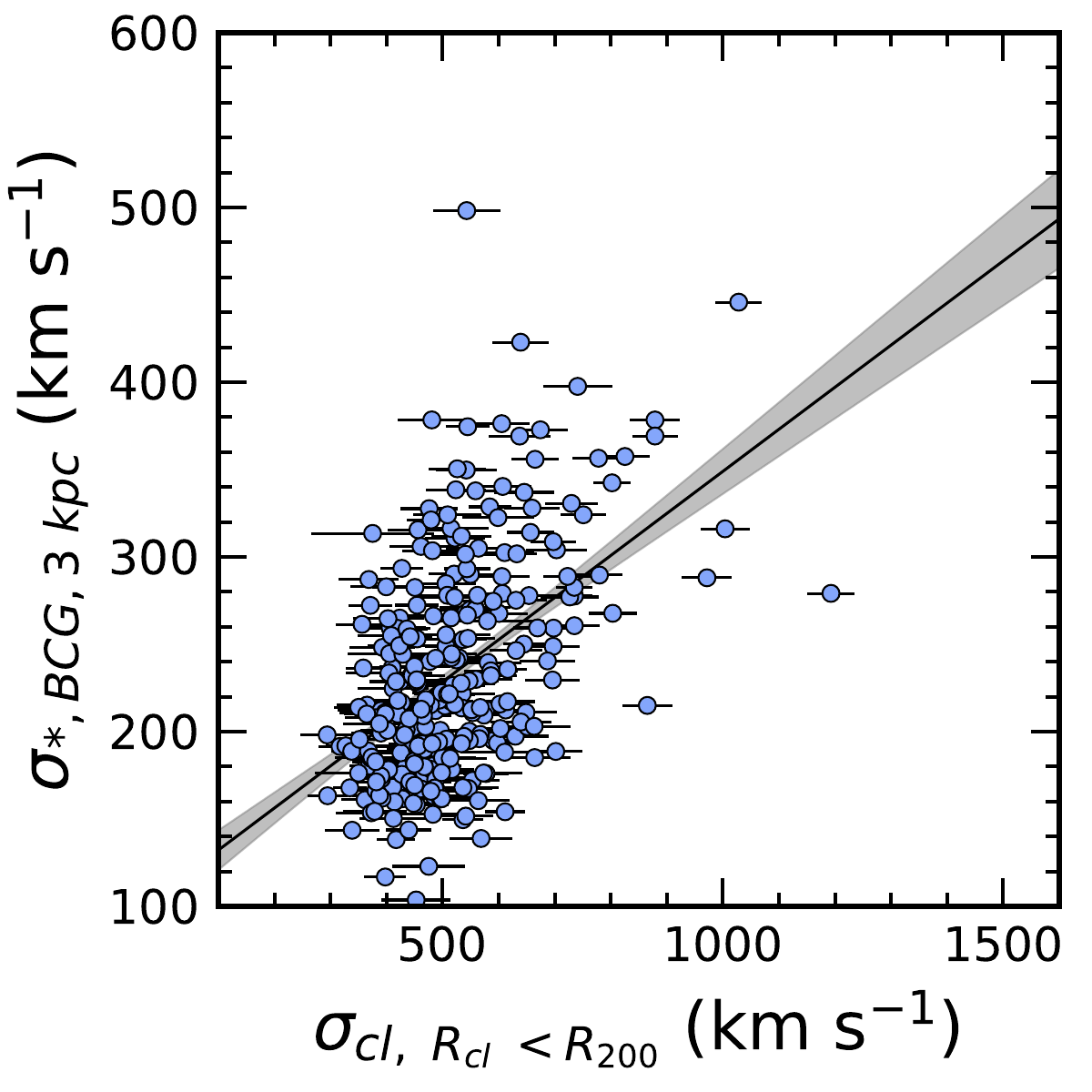}
\caption{The BCG velocity dispersion as a function of cluster velocity dispersion. The solid line shows the best-fit relation. Gray shaded region shows the $1\sigma$ distribution of the MCMC fits. }
\label{sigma_sigma}
\end{figure}

Figure \ref{sigma_sigma} compares $\sigma_{cl}$ and $\sigma_{*, BCG}$. In general, higher $\sigma_{cl}$ clusters tend to host BCGs with large $\sigma_{*, BCG}$, but the scatter is large at low $\sigma_{cl}$. The Spearman rank-order correlation test yields a correlation coefficient of 0.45 with a $p-$value $1.56 \times 10^{-15}$. The Pearson correlation test yields a correlation coefficient of 0.49 with $p-$value $2.29 \times 10^{-18}$. We also derive the best-fit relation between $\sigma_{cl}$ and $\sigma_{*, BCG}$: 
$\sigma_{*, BCG} = (0.295 \pm 0.049)~ \sigma_{cl} + (95 \pm 25).$ 

\section{Discussion}\label{sec:discuss}

We derive a relation between the velocity dispersions of cluster halos in the TNG300-1 simulation and their BCGs. The velocity dispersions of clusters and their BCGs probe the masses of both the clusters and their BCGs. Here, we compare the simulated relation in Section \ref{sec:comp} with the data (Section \ref{sec:obs}). We also explore the simulated redshift evolution of this relation in Section \ref{sec:redshift}. 

\subsection{Comparison with HeCS-omnibus}\label{sec:comp}

We compare the correlation between $\sigma_{cl}$ and $\sigma_{*, BCG}$ derived from the TNG300-1 halos with the observed relation. We describe the observational sample in Section \ref{sec:obs}. 

\begin{figure*}
\centering
\includegraphics[scale=0.47]{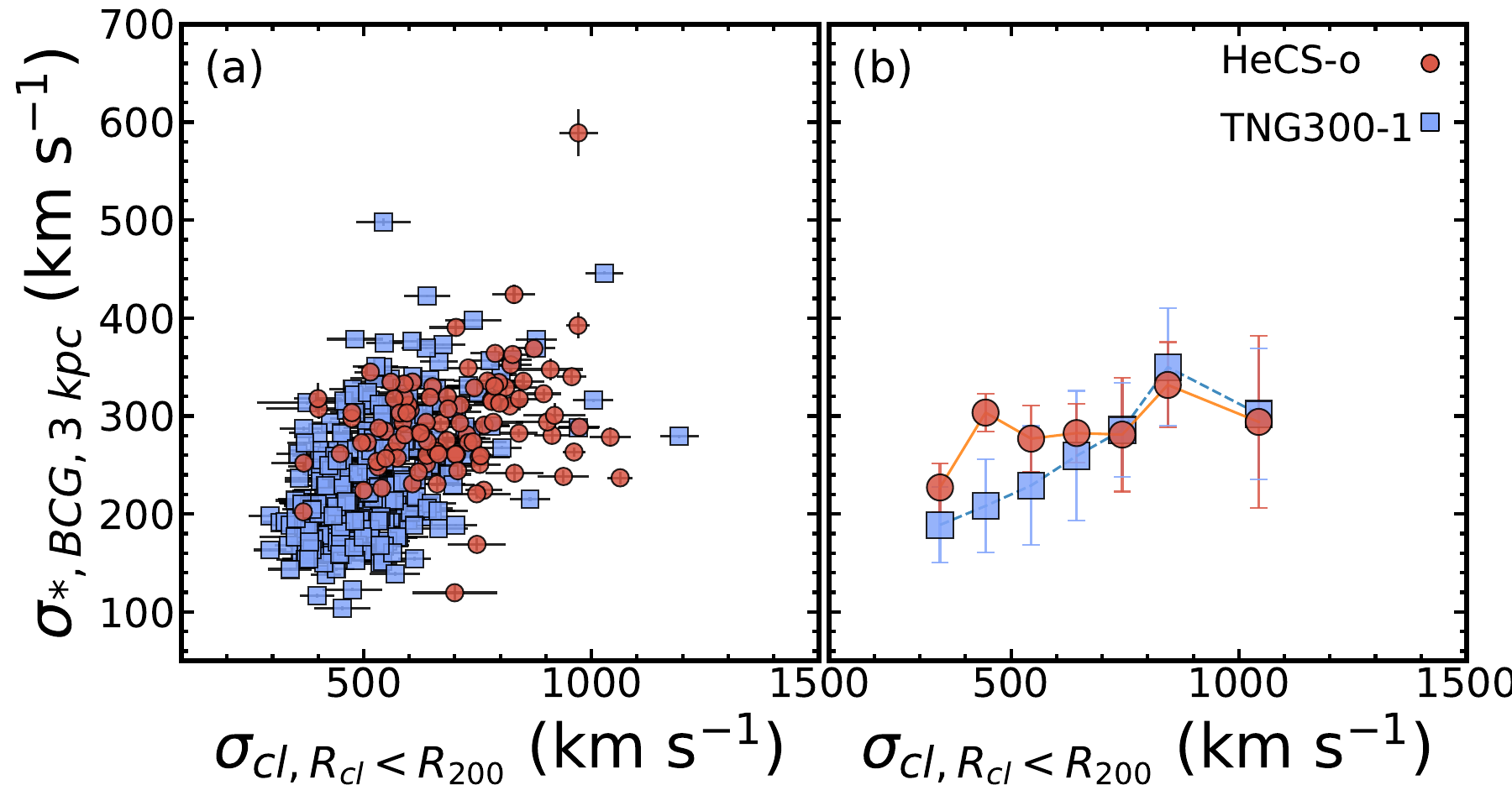}
\caption{(a) $\sigma_{*, BCG, 3 kpc}$ vs. $\sigma_{cl}$ distribution of TNG clusters (blue squares) and HeCS-omnibus clusters (red circles). (b) The median distribution. The error bar indicates $1\sigma$ standard deviation. }
\label{comp_sigma_sigma}
\end{figure*}

Figure \ref{comp_sigma_sigma} (a) shows the $\sigma_{*, BCG} - \sigma_{cl}$ relation for simulated clusters (blue squares) and for the observed HeCS-omnibus subsample (red circles, Section \ref{sec:obs}). The observed clusters overlap the simulated cluster sample. A difference between the two samples is the lack of observed clusters with small $\sigma_{BCGs}$ at low $\sigma_{cl}$ (i.e., $\sigma_{cl} \lesssim 600~\kms$). Figure \ref{comp_sigma_sigma} (b) demonstrates this difference; Figure \ref{comp_sigma_sigma} (b) displays the median $\sigma_{*, BCG}$ in various $\sigma_{cl}$ bins. The median $\sigma_{*, BCG}$ of the simulated clusters at $\sigma_{cl} \lesssim 600~\kms$ is smaller than for the observed clusters; however, this difference is less than 2$\sigma$. 

\begin{figure*}
\centering
\includegraphics[scale=0.47]{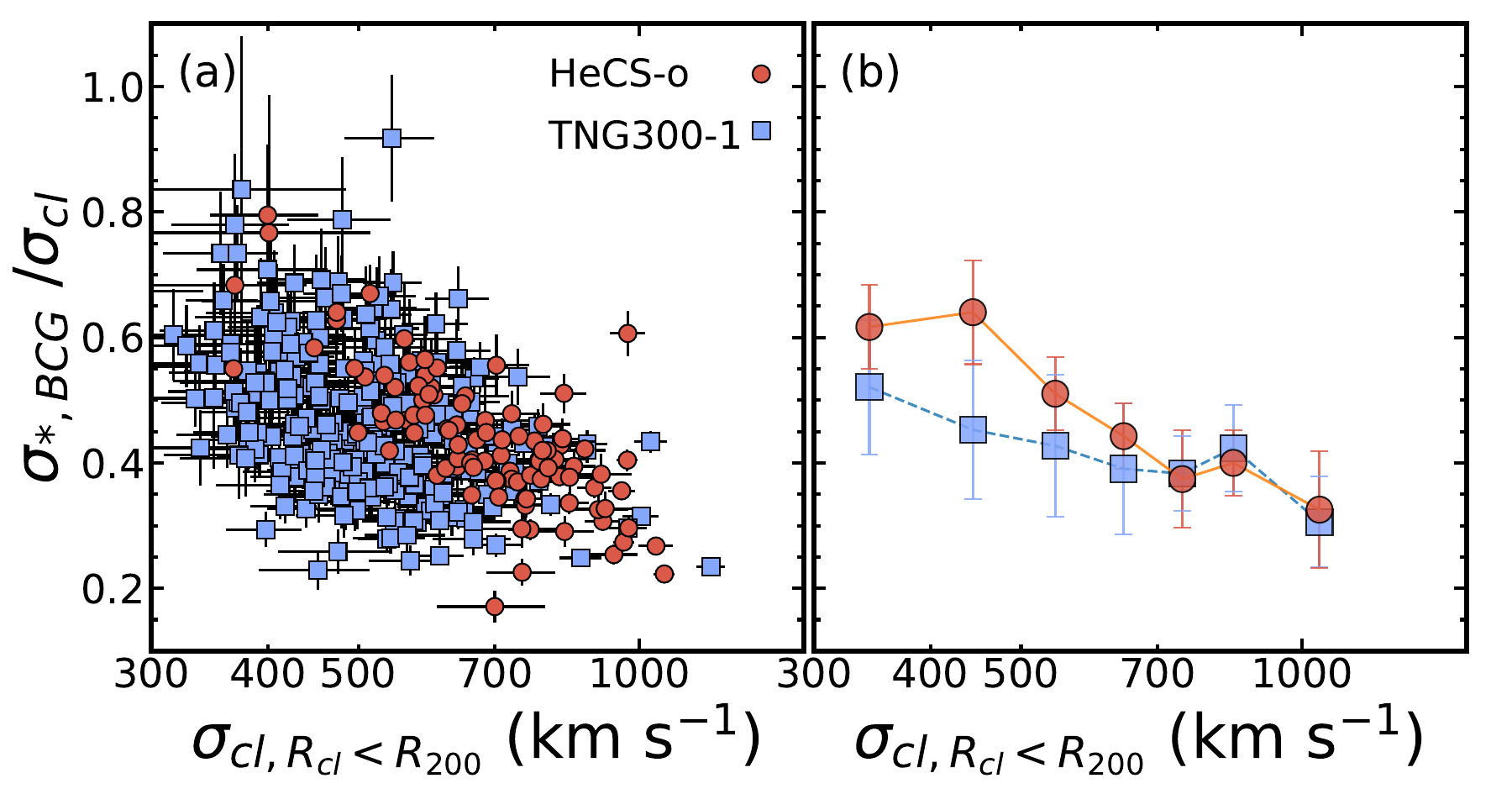}
\caption{(a) The ratio between $\sigma_{*, BCG, 3 kpc}$ and $\sigma_{cl}$ as a function of $\sigma_{cl}$. The $x-$axis is in log scale. (b) The median distribution of $\sigma_{*, BCG, 3 kpc} / \sigma_{cl}$ ratio as a function of $\sigma_{cl}$.}
\label{comp_sigma_ratio}
\end{figure*}

Figure \ref{comp_sigma_ratio} (a) displays the ratio between $\sigma_{*, BCG} / \sigma_{cl}$ as a function of $\sigma_{cl}$. The observed clusters show a very tight relation: the ratio decreases as the cluster mass increases. Many simulated clusters overlap the observed clusters. However, there are a more  clusters with a low $\sigma_{*, BCG} / \sigma_{cl}$ ratio at $\sigma_{cl} \lesssim 600~\kms$. These systems are absent in the observed sample. Figure \ref{comp_sigma_ratio}(b) highlights this difference: the median ratio between the observed and simulated clusters differs slightly for $\sigma_{cl} \lesssim 600~\kms$. 

There are several interesting aspects in these comparisons. First, the trend we derive from the TNG300-1 differs slightly from previous simulations. \citet{Dolag10} and \citet{Remus17} predict a constant ($\sigma_{*, BCG}/\sigma_{cl} \sim 0.5)$ over a large $\sigma_{cl}$ range. However, for a large fraction of the TNG300-1 clusters, $\sigma_{*, BCG} / \sigma_{cl}$ decreases as $\sigma_{cl}$ increases. Furthermore, the median $\sigma_{*, BCG} / \sigma_{cl}$ is below the ratio predicted by the earlier simulations. 

The result from TNG300-1 is consistent with more recent simulations. \citet{Marini21} use the DIANOGA set of hydrodynamically simulated clusters to explore the $\sigma_{*, BCG} - \sigma_{cl}$ relation. In their simulation, \citet{Marini21} show that the $\sigma_{*, BCG}$ versus $M_{200}$ relation is shallower than previously reported by simulations \citet{Dolag10} and \citet{Remus17}. Consequently, the $\sigma_{*, BCG} - \sigma_{cl}$ derived by \citet{Marini21} is consistent with the observed relation.

\subsection{Redshift Evolution of the $\sigma_{*, BCG} - \sigma_{cl}$ Relation}\label{sec:redshift}

Simulations like TNG300-1 enable exploration of the evolution of the $\sigma_{*, BCG} - \sigma_{cl}$ relation at different redshifts. There are two approaches for exploring redshift evolution. First, we can trace the history of the most massive clusters and their BCGs identified in the current universe. For example, we select 346 clusters and their BCGs in the current universe for deriving the $\sigma_{*, BCG} - \sigma_{cl}$ relation. By measuring the properties of their progenitors at higher redshift, we can probe the redshift evolution of the correlation between cluster and BCGs. \citet{Marini21} explore this relation for $0 < z < 2$. We plan to explore this  evolution using TNG300-1 in a companion paper (J.Sohn et al. 2022, in preparation). 

Here, we take an approach more directly tied to observations for exploring the redshift evolution of $\sigma_{*, BCG} - \sigma_{cl}$ relation. We first select the most massive cluster halos at different redshifts. We then compare the cluster and BCG properties.  

We select massive halos with $M_{200} > 10^{14}$ M$_{\odot}$ at six different redshift snapshots (i.e., $z = 0.0, 0.1, 0.2, 0.5, 0.7$ and, 1.0). In each snapshot, there are 280, 250, 230, 149, 107, and 50 halos with $M_{200} > 10^{14}$ M$_{\odot}$. We apply the same method that we used for the $z = 0$ sample to derive the velocity dispersion of clusters and their BCGs.

\begin{figure*}
\centering
\includegraphics[scale=0.40]{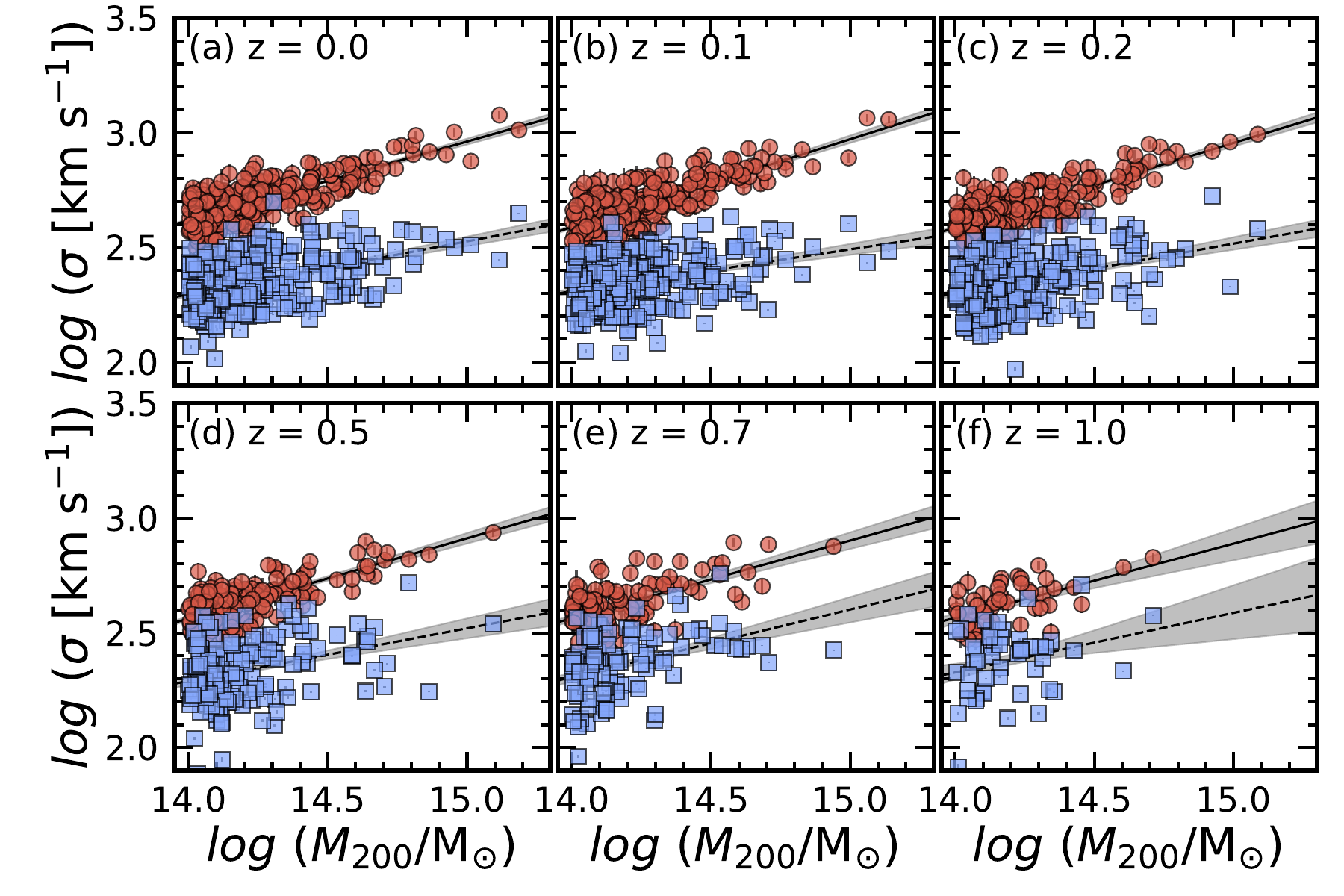}
\caption{Same as Figure \ref{sigma_sigma_m200}, but for six different redshift snapshots: (a) $z = 0.0$, (b) $z = 0.1$, (c) $z = 0.2$, (d) $z = 0.5$, (e) $z = 0.7$, and (f) $z = 1.0$. The solid and the dashed lines show the best-fit relations. Gray shaded regions show the $1\sigma$ distributions of the MCMC fits. }
\label{sigma_sigma_m200_redshift}
\end{figure*}

\begin{figure*}
\centering
\includegraphics[scale=0.40]{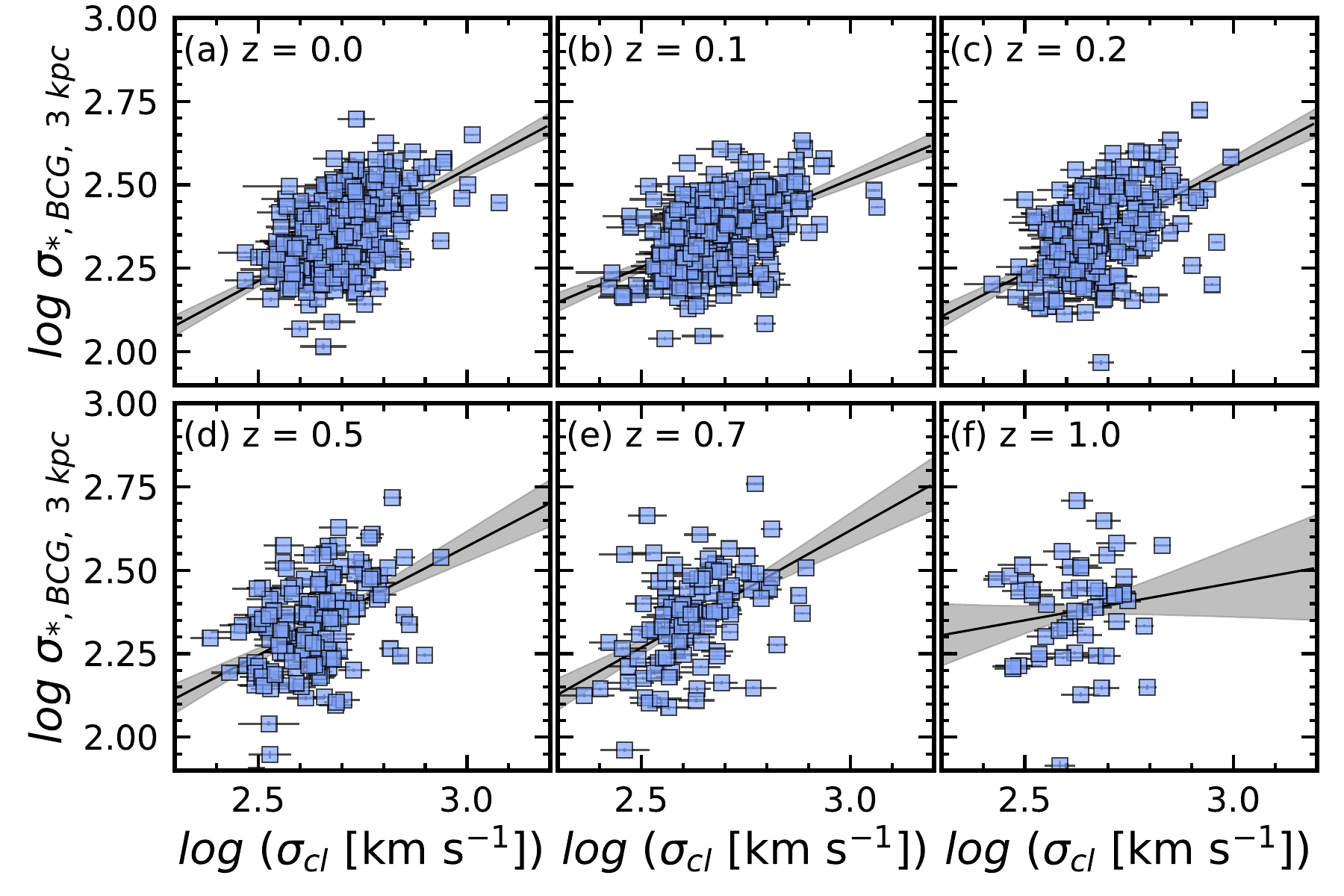}
\caption{Same as Figure \ref{sigma_sigma_m200}, but for the relation derived from six different redshift snapshots: (a) $z = 0.0$, (b) $z = 0.1$, (c) $z = 0.2$, (d) $z = 0.5$, (e) $z = 0.7$, and (f) $z = 1.0$. Gray shaded regions mark $1\sigma$ distribution of the MCMC fits. Black solid lines show the best-fit relations. }
\label{bsigma_csigma_redshift}
\end{figure*}

Figure \ref{sigma_sigma_m200_redshift} shows $\sigma_{cl}$ (red circles) and $\sigma_{*, BCG}$ (blue squares) as a function of cluster mass ($M_{200}$) at six different redshifts. $\sigma_{cl}$ is tightly correlated with the $M_{200}$ over the redshift range. However, the correlation between $\sigma_{*, BCG}$ and $M_{200}$ is less tight at higher redshift. 

We plot the $\sigma_{*, BCG} - \sigma_{cl}$ relation at six different redshifts (Figure \ref{bsigma_csigma_redshift}). The solid lines show the best-fit linear relations based on the MCMC technique. Table \ref{tab:ssevol} summarizes the slopes of these best-fit relations; the slope and its uncertainty correspond to the median and $1\sigma$ standard deviation of the MCMC fits. 

There are two notable changes in the relation as a function of redshift. First, the range of $\sigma_{cl}$ broadens as universe ages. The $\sigma_{cl}$ of $M_{200} > 10^{14}$ M$_{\odot}$ clusters at $z = 1$ ranges from $270~\kms$ to $674~\kms$; at $z = 0$ the range is $294~\kms$ to $1192~\kms$. The maximum $\sigma_{cl}$ increases by $\sim 70\%$.

Second, in contrast to the $\sigma_{cl}$ ranges, the range of $\sigma_{*, BCG}$ narrows at lower redshifts: $82 < \sigma_{*, BCG}~(\kms) < 510$ at $z = 1.0$ to $103 < \sigma_{*, BCG}~(\kms) < 498$ at $z = 0.0$. Additionally, the $\sigma_{*, BCG}$ of a cluster with a similar $\sigma_{cl}$ at higher redshift is generally larger than for its counterpart at lower redshift. As a consequence, the slope of the $\sigma_{*, BCG} - \sigma_{cl}$ relation changes, although the uncertainty is large. 

A decreasing $\sigma_{*, BCG}$ over time appears inconsistent with the general idea that the mass of BCG subhalo ($\propto~ \sigma_{*, BCG}$) grows via accretion and mergers as the universe ages. However, the $\sigma_{*, BCG}$ measurement is sensitive to the dynamical stage of the BCGs, particularly when the BCGs undergo active interactions with other galaxies (Figure \ref{bcg_rv}). The interactions between BCGs and other cluster members occur more often at higher redshift, where BCG are actively forming. 

\begin{figure*}[ht]
\centering
\includegraphics[scale=0.24]{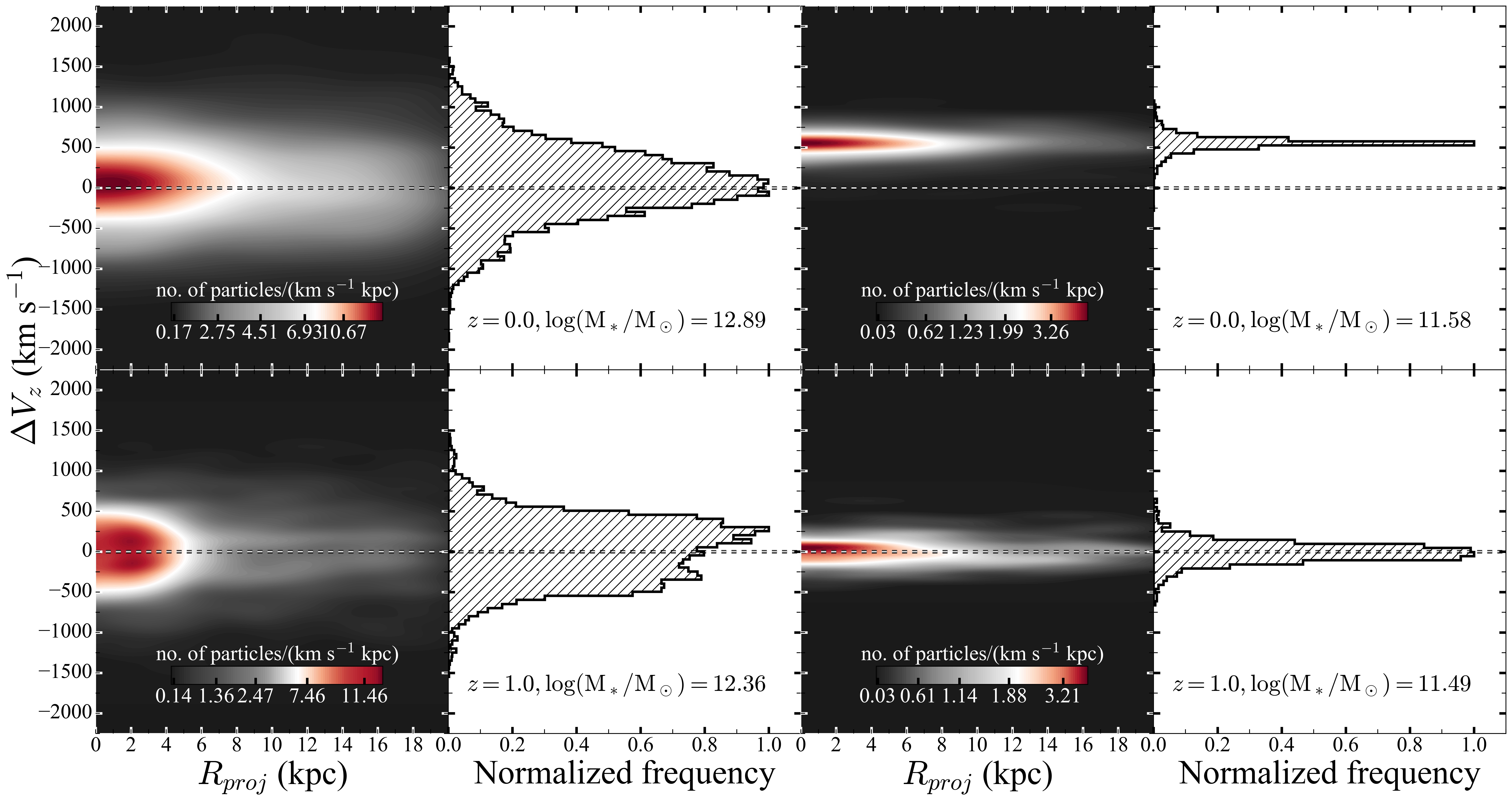}
\caption{R-v diagrams (left) and line-of-sight velocity distributions of stellar particles within BCGs at $z = 0$ (top panels) and at $z = 1$ (bottom panels). The BCGs at each redshift lie within the most massive (left two panels) and least massive (right two panels) clusters. }
\label{BCG_rv_redshift}
\end{figure*}

Figure \ref{BCG_rv_redshift} demonstrates the dynamically unrelaxed nature of the BCGs at higher redshift. The upper panels of Figure \ref{BCG_rv_redshift} display the R-v diagrams of stellar particles within two BCGs at $z = 0$. These BCGs are the most massive subhalo in the most massive (left) and least massive (right) clusters. We also show the line-of-sight velocity distributions of the stellar particles within $R_{proj} < 1$ kpc in the panels adjacent to the R-v diagrams. The lower panels show similar BCGs at $z = 1$. 

The difference between the R-v diagrams of the most massive BCGs at $z = 0$ and $z = 1$ (left panels) is dramatic. The BCG at $z = 0$ shows well defined trumpet-like patterns in the core and an extended distribution in the outer region ($R_{proj} > 5$ kpc). In contrast, the most massive BCG at $z = 1$ shows complex structure in the core ($R_{proj} < 1$ kpc, line-of-sight velocity distribution). This structure presumably inflates the velocity dispersion. Less massive BCGs at $z = 1$ also show elongated stellar distributions along the line of sight. These elongations exceed those of BCGs with similar mass at $z = 0.0$. Furthermore, the high-redshift BCGs show more clumpy structures in the outer region.

The redshift evolution of the $\sigma_{*, BCG} - \sigma_{cl}$ relation offers a unique test-bed for observational studies of cluster and BCG formation. The next generation spectroscopic surveys, including DESI, 4MOST, and Subaru Prime Focus Spectrograph (PFS), will increase spectroscopic samples of clusters at $z < 1$. Then, a large number of clusters at high redshift with velocity dispersion measurements based on large numbers of members will be available. The central stellar velocity dispersions of the BCGs in these high redshift clusters will also be measured.

\begin{deluxetable}{llcccc}
\label{tab:ssevol}
\tablecaption{$\sigma_{*, BCG} - \sigma_{cl}$ relation$^{*}$ at different redshift}
\tablecolumns{3}
\tabletypesize{\scriptsize}
\tablewidth{0pt}
\tablehead{
\colhead{redshift} & \colhead{$\alpha$} & \colhead{$\beta$}}
\startdata
0.0 & $ 0.67 \pm 0.07$ & $0.54 \pm 0.20$ \\
0.1 & $ 0.53 \pm 0.07$ & $0.94 \pm 0.19$ \\
0.2 & $ 0.65 \pm 0.08$ & $0.62 \pm 0.22$ \\
0.5 & $ 0.65 \pm 0.13$ & $0.62 \pm 0.33$ \\
0.7 & $ 0.70 \pm 0.14$ & $0.51 \pm 0.36$ \\
1.0 & $ 0.23 \pm 0.28$ & $1.79 \pm 0.72$
\enddata 
\tablenotetext{*}{$\log \sigma_{*, BCG} = \alpha~\log \sigma_{cl} + \beta$}
\end{deluxetable}

\section{Conclusion} \label{sec:conclusion}

We explore the $\sigma_{*, BCG} - \sigma_{cl}$ scaling relation based on the Illustris-TNG simulations. TNG300-1 includes 280 massive clusters with $M_{200} > 10^{14}$ M$_{\odot}$. We measure the line-of-sight velocity dispersion of stellar particles within 3 kpc of the BCG center as $\sigma_{*, BCG}$. We compute the line-of-sight velocity dispersion of cluster members within $R_{cl} < R_{200}$ as $\sigma_{cl}$. These $\sigma_{cl}$ and $\sigma_{*, BCG}$ correspond to properties of observed clusters. 

Both $\sigma_{*, BCG}$ and $\sigma_{cl}$ of the simulated clusters are correlated with the halo mass ($M_{200}$). The $\sigma_{*, BCG} - M_{200}$ scaling relation shows a larger scatter than the $\sigma_{cl} - M_{200}$ scaling relation. The slopes of the relation differ: the $\sigma_{*, BCG} - M_{200}$ relation is slightly shallower than the $\sigma_{cl} - M_{200}$ relation. This difference in slope produces a $\sigma_{*, BCG} - \sigma_{cl}$ relation with a slope that departs from unity.

We compare the $\sigma_{*, BCG} - \sigma_{cl}$ scaling relation with that derived from the HeCS-omnibus sample. Overall, the simulated and observed scaling relations overlap. In the simulation, there are many clusters with low $\sigma_{cl}$ ($< 600~\kms$) and low $\sigma_{*, BCG}$ ($< 200~\kms$). These systems are absent in HeCS-omnibus. The lack of these systems presumably result from selection effects in HeCS-omnibus; less massive clusters with less massive BCGs are excluded at least in part as a result of selection in X-ray luminosity. The low $\sigma_{cl}$ clusters make the simulated scaling relation slightly steeper than the observed relation, but the difference is within the uncertainty. 

We trace the $\sigma_{*, BCG} - \sigma_{cl}$ scaling relations as a function of redshift. We select massive cluster halos with $M_{200} > 10^{14}$ M$_{\odot}$ at six different redshifts $\lesssim 1$. We sample the simulations in a way that provides a direct testbed for future high-redshift observations of these scaling relation. 

The $\sigma_{*, BCG}$ is correlated with $\sigma_{cl}$ over the redshift range ($z < 1.0$) we explore. However, the correlation is weaker for massive clusters at higher redshift because of the large scatter in $\sigma_{*, BCG}$s. The BCGs in higher redshift massive cluster are actively interacting with other cluster members inflating the velocity dispersion. Future observations of high redshift clusters and their BCGs will be a clean test of these models. 

\acknowledgements
We thank Josh Borrow for carefully reading the draft and for providing valuable comments that improved the paper. We also thank Antonaldo Diaferio and Ken Rines for insightful discussions. J.S. is supported by the CfA Fellowship. M.J.G. acknowledges the Smithsonian Institution for support. MV acknowledges support through NASA ATP 19-ATP19-0019, 19-ATP19-0020, 19-ATP19-0167, and NSF grants AST-1814053, AST-1814259, AST-1909831, AST-2007355 and AST-2107724. MV also acknowledges support from a MIT RSC award, the Alfred P. Sloan Foundation, and by NASA ATP grant NNX17AG29G. I.D. acknowledges the support of the Canada Research Chair Program and the Natural Sciences and Engineering Research Council of Canada (NSERC, funding reference number RGPIN-2018-05425).

All of the primary TNG simulations have been run on the Cray XC40 Hazel Hen supercomputer at the High Performance Computing Center Stuttgart (HLRS) in Germany. They have been made possible by the Gauss Centre for Supercomputing (GCS) large-scale project proposals GCS-ILLU and GCS-DWAR. GCS is the alliance of the three national supercomputing centres HLRS (Universitaet Stuttgart), JSC (Forschungszentrum Julich), and LRZ (Bayerische Akademie der Wissenschaften), funded by the German Federal Ministry of Education and Research (BMBF) and the German State Ministries for Research of Baden-Wuerttemberg (MWK), Bayern (StMWFK) and Nordrhein-Westfalen (MIWF). Further simulations were run on the Hydra and Draco supercomputers at the Max Planck Computing and Data Facility (MPCDF, formerly known as RZG) in Garching near Munich, in addition to the Magny system at HITS in Heidelberg. Additional computations were carried out on the Odyssey2 system supported by the FAS Division of Science, Research Computing Group at Harvard University, and the Stampede supercomputer at the Texas Advanced Computing Center through the XSEDE project AST140063.

\bibliographystyle{aasjournal}
\bibliography{ms}

\end{document}